\documentclass[apjl]{emulateapj}
\usepackage{apjfonts}

\slugcomment{submitted to ApJL}

\shorttitle{The Argo Star System}

\shortauthors{Rocha-Pinto et al.}

\begin{document}


\title{The Dog on the Ship: The ``Canis Major Dwarf Galaxy'' \\ as an
  Outlying Part of the Argo Star System}


\author{
H. J. Rocha-Pinto\altaffilmark{1,2},
S. R. Majewski\altaffilmark{2},
M. F. Skrutskie\altaffilmark{2},
R. J. Patterson\altaffilmark{2},
H. Nakanishi\altaffilmark{3},
R. R. Mu\~noz\altaffilmark{2}, \&
Y. Sofue\altaffilmark{4}
}

\altaffiltext{1}{Observat\'orio do Valongo, UFRJ, Rio de Janeiro, RJ
  20080-090, Brazil (helio@ov.ufrj.br)}

\altaffiltext{2}{Department of Astronomy, University of Virginia,
    Charlottesville, VA 22903 (srm4n,mfs4n,rjp0i,rrm8f@virginia.edu)}

\altaffiltext{3}{Nobeyama Radio Observatory, Minamimaki, Minamisaku,
    Nagano 384-1305 (hnakanis@nro.nao.ac.jp)}

\altaffiltext{4}{Institute of Astronomy, The University of Tokyo, 2-21-1 Osawa,
    Mitaka, Tokyo 181-0015}

\begin{abstract}

  Overdensities in the distribution of low latitude, 2MASS giant stars are
  revealed by systematically peeling away from sky maps the bulk of the giant
  stars conforming to ``isotropic'' density laws generally
  accounting for known Milky Way components.
  This procedure, combined with a higher resolution treatment of the sky density
  of both giants and dust allows us to probe to lower Galactic
  latitudes than previous 2MASS giant star studies.  While the results show
  the swath of excess giants previously associated with the Monoceros
  ring system in the second and third Galactic quadrants at distances of
  $6-20$ kpc, we also find a several times
  larger overdensity of giants in the same distance range concentrated
  in the direction of the ancient constellation Argo.
  Isodensity contours of the large structure suggest that it is highly
  elongated and inclined by about $3\arcdeg$ to the disk, although
  details of the structure --- including the actual location of highest
  density, overall extent, true shape --- and its origin, remain
  unknown because only a fraction of it lies outside highly
  dust-obscured, low latitude regions.  Nevertheless, our results suggest
  that the 2MASS M giant overdensity previously claimed to represent
  the core of a dwarf galaxy in Canis Major ($l\sim240\arcdeg$) is
  an artifact of a dust extinction window opening to the overall density rise
  to the more significant Argo structure centered at larger
  longitude ($l\sim 290\arcdeg\pm 10\arcdeg$, $b\sim -4\arcdeg\pm 2\arcdeg$).

\end{abstract}


\section{Introduction}

\citet{newberg} identified a low-$b$ excess of stars in the
Sloan Digital Sky Survey (SDSS) at $[l,b]\sim [223,+20]\arcdeg$,
with distances beyond, and with a larger apparent
scaleheight than, the nominal Milky Way (MW) disk.  They attributed this
excess to the presence of either a newly discovered dwarf galaxy or
tidal stream lying just outside the Galactic disk.
\citet[][M03 hereafter]{maj03} identified the same structure 
in their study of 2MASS M giants.  \citet{ibata} observed the
structure with optical CMDs for fields spanning over $100\arcdeg$ of
longitude, and proposed as a most likely explanation of this apparent
``ring'' around the MW disk that it is ``a perturbation of the disk,
possibly the result of ancient warps''.  Meanwhile, \citet{yanny} relied
on SDSS spectra and imaging in other areas to argue the structure
represents the remains of a tidally disrupted MW satellite, whereas
\citet{helmi} explored models of the structure as either a ``transitory
localized radial density enhancement'' from particles stripped off a
satellite on a recent peri-Galactic passage, or the analogue of
``shells'' found around elliptical galaxies deriving from minor mergers
several Gyr in the past.

More detailed study of the distribution of 2MASS M giants has shown that
(1) this newly-found structure is indeed ring-like, spanning $>170\arcdeg$ around the
disk at a typical Galactocentric distance of $\sim16$ kpc and broadened
into both north and south Galactic hemispheres (\citealt{PaperI},
\citeauthor*{PaperI} hereafter; \citealt{martin04a}, \citeauthor*{martin04a}
hereafter), (2) the structure contains stars enriched $>10\times$ higher
than the [Fe/H]$=-1.6$ previously reported by \citet{yanny}, belying
a large metallicity spread
(\citeauthor*{PaperI}; \citealt{crane}) like that seen in MW satellite galaxies, and (3)
the stars in the structure follow a slightly non-circular orbit with
a relatively low ($13-20\,{\rm km\,s^{-1}}$) velocity dispersion
\citep{crane,martin04b}.
Both globular {\it and} old open clusters spatially and kinematically
correlated to the structure have been identified
\citep{crane,frinchaboy,bella}; these clusters
follow a well-defined age-metallicity relation.  This combined evidence
supports the view that the anticenter ``ring´´ represents tidal debris from
a Sagittarius (Sgr) dwarf-like galaxy along an orbit with low
inclination to, and orbital radius just larger than, the MW disk.

More recently, probing asymmetries in the 2MASS M giant distribution
about the Galactic plane, \citeauthor*{martin04a} claim to have found the core
of a satellite galaxy in Canis Major (``CMa''; $l \sim 240\arcdeg$), which they
argue to be the progenitor of the ring.
However, \citet{momany} have argued that CMa may be the MW disk stellar warp, based on
comparison of observed to synthetic CMDs generated by models of a
warped MW disk. This has prompted vigorous
debate over the nature of CMa \citep{bella,martin04b,sbordone,penarrubia}.
Whatever its origin, that there is a ``CMa overdensity'' has
apparently gained wide acceptance \citep[e.g.,][]{forbes, tosi,
  sbordone, martinez}.

In this contribution we call into question a ``core'' overdensity in CMa,
and find an apparently more significant
giant star overdensity at $l\sim 290\arcdeg$. 

\section{Data and Comparison Model}

We extract a giant star candidate sample from the 2MASS all-sky point source
release as all stars having dereddened $0.85 < J-K_S < 1.5$, $K_S < 13.0$,
photometric quality flag {\tt AAA}, 
and that meet the M giant color locus restriction used by M03.
For each star we derive a distance probability distribution function
(DPDF) based on its dereddened, PSF-fitted $K_S$ and $J-K_S$,
the variation of the color-absolute magnitude relation with [Fe/H]
\citep{ivanov}, and an assumed [Fe/H] distribution for
giants typical of the ring (a Gaussian
with mean [Fe/H]$=-1$ and spread $0.4$ dex).  Calculation of the
DPDF is described in more detail in \citeauthor*{PaperI}.
We adopt as a single distance estimate, $r$, for each star the mode of its DPDF.

Extinctions were calculated as
$(A_J,A_{K_S},E_{J-K_S}) = (0.90, 0.25, 0.65) \alpha E_{B-V}$, where
$E_{B-V}$ is from \citet*{schlegel} and $\alpha=\case{2}{3}$ \citep{triand}.
As with previous, similar studies (\citeauthor*{PaperI, martin04a}) our analysis
is limited to $E_{B-V} \le 0.55$ to avoid false overdensity signals
from small-scale differential reddening or errors in its
estimation. However, we make several improvements
over previous work:
(1) To more closely follow the $E_{B-V} = 0.55$ ``border''
our maps are binned into finer, $1\arcdeg\times 1\arcdeg$
$(l,b)$ bins than the $4\arcdeg \times 2\arcdeg$ bins of
\citeauthor*{martin04a} or the $4\arcdeg\times 4\arcdeg$ bins of
\citeauthor*{PaperI}.
(2) \citeauthor*{martin04a} depended on direct comparisons of one region of the sky
to another presumed to be ``symmetric'' --- a procedure that requires
{\it both} regions to have $E_{B-V} \le 0.55$ and that ultimately removes
large regions of ``good'' sky from consideration because comparison
fields are too reddened.
Here we rely on a Galactic model of
the number of stars from canonical Galactic populations
as a function of distance and $(l,b)$
as a reference for all $E_{B-V} \le 0.55$ fields, which allows
us to probe areas of the sky previously ignored.

The densities of the asymmetries sought are small compared to the MW
foreground density.  The goal of creating a Galactic model is to
``fairly'' remove the majority of this foreground to highlight the
residual.  Rocha-Pinto et al.\ (2006; in preparation)
describes the Galaxy model in detail; here we
outline the general procedures used to construct it.  The key to such a
model is that it be smooth, have a cylindrically symmetric density
distribution about the Galactic center, and be constrained globally by
the observed density distribution in available, low $E_{B-V}$ lines of
sight, so that, in the mean, most of the 2MASS giants are removed when the
model is subtracted from the ``observed'' $(l,b,r)$ distribution to
leave behind a high contrast picture of ``overdensities''.  To this end, the
model does not need to be exact, nor should it necessarily be
interpreted as a true description of the MW giant star
distribution, though we start from a parameterization meant to produce a
classical ``starcount model'' from thin disk, thick disk and halo
density laws \citep[those of][]{siegel}.
These density laws are multiplied by adopted luminosity functions for the
corresponding stellar populations and integrated from
the brightest to faintest absolute magnitudes of stars that can be in
our observed sample.  After integration over solid angle ($1\arcdeg\times
1\arcdeg$) and normalization of the results to the observed stellar density
at longitudes not contaminated by known stellar
streams or the Magellanic Clouds, we have the expected number of stars
coming from each contributing MW population within a given
$r$ range.

\begin{figure}
\includegraphics[scale=0.9]{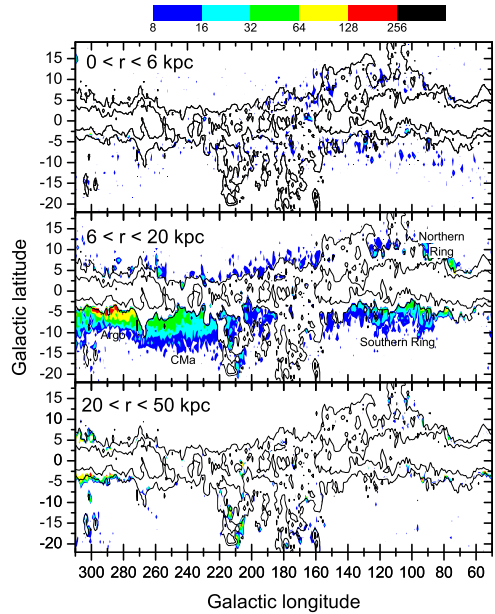}
\caption{Residual 2MASS late type giant star density with respect
  to our MW model projected onto the celestial sphere at different
  distance ranges. Contour levels represent geometrically increasing densities 
  in number of stars deg$^{-2}$.  The solid lines show
  $E_{B-V}$ contour levels at 0.55 and 1.20 mag; regions
  having $E_{B-V} > 0.55$ have been discarded from the analysis.  This figure
  supports our claim that the ``CMa´´ overdensity is only an outlying part of the
  larger Argo overdensity seen in the $6 < r \le 20$ kpc range.
\label{argo}}
\end{figure}

Because the 2MASS stars have distances calculated as if they have come
from a hypothetical stellar population with $\langle{\rm [Fe/H]}\rangle
= -1.0$ and $\sigma_{\rm [Fe/H]}$= 0.4, stars from the simulated MW have their
distances {\it recalculated} with a DPDF in the same way --- i.e. as if
their simulated metallicity were unknown.  That our model MW succeeds in
removing the bulk of the canonical MW populations of giants is shown in
the difference between the observed and model 3D density distributions
(Fig.\ \ref{argo}), where stellar densities in the projected distance
ranges $r \le 6$ kpc and $20 < r \le 50$ kpc are essentially eliminated.
In contrast, in the distance range dominated by the ring, $6
< r \le 20$ kpc, significant residual densities --- which we claim are true
{\it overdensities} --- can be seen.

\section{The Argo Overdensity}

Most of the ``$\backsim$''-shaped arc of overdensity sweeping all of the way
across the middle panel of Figure 1 can be attributed to the ``anticenter
ring/warp/Monoceros stellar structure,'' discussed by the various
studies referenced in \S1.
Here we focus specifically on the very obvious, large overdensity in the third and
fourth quadrants from about $l=210\arcdeg$
to at least the left edge ($l=310\arcdeg$) of our sample range and
peaking at $[l,b] \sim [290,-4]\arcdeg$.  From
the overall rise in density towards this specific point in Figure 1, we
surmise that the trend continues and an even higher
overdensity peak may lie somewhere in the obscured region with $-4\arcdeg
< b < +2\arcdeg$ near $285\arcdeg < l < 300\arcdeg$, which
spans the constellations Carina, Vela and Puppis.
Because the peak of this large overdensity remains uncertain and possibly obscured, we
call the structure ``Argo,'' after the large, ancient 
constellation later dismembered into Carina, Vela and Puppis by the IAU.
As may be seen, the density of
this region is significantly higher than any other visible giant star
excesses --- e.g., it is 4 to 8 times more dense than the densest,
visible parts of the Northern and Southern Ring in the second
quadrant.

\begin{figure}
\plotone{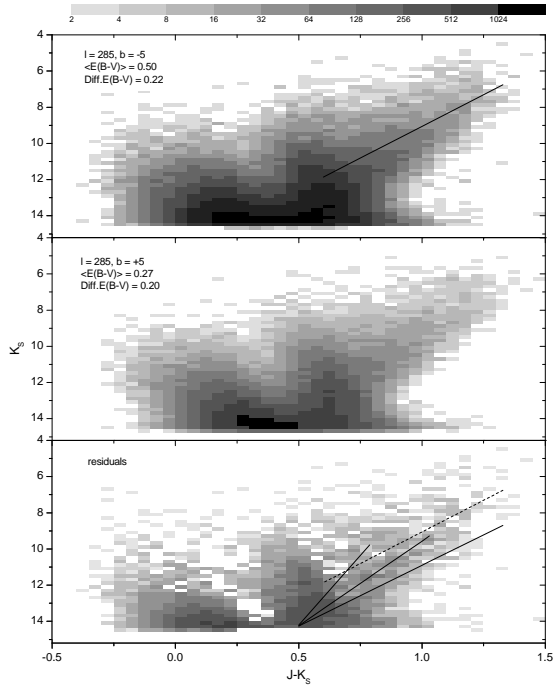}
\caption{2MASS CMDs for (top) a field with an excess of stars
  associated with the Argo overdensity, (middle) a control field (same
  $l$, opposite $b$), and (bottom) their difference.  The line in the top panel shows that
  the bulk of MW giants can be fitted by an [Fe/H]$=-0.1$ RGB isochrone having a
  median distance modulus
  of 14 mag (i.e. 6.3 kpc).  This isochrone is reproduced in the
  lower panel as a dashed line to demonstrate that the residual RGB we associate with
  Argo is mostly fainter.
  The solid lines in the bottom panel show
  [Fe/H]$=-1.4$, $-0.7$ and $+0.0$ RGB isochrones placed 13.8 kpc from the Sun
  to match the distinct RGB from Argo as well as the
  previously suggested (see \S1) large [Fe/H] spread of the anticenter ring.
  Residue at $J-K_s \sim 0.5$ may represent Argo red clump stars.
  \label{hess}}
\end{figure}

The overall impression is that this large excess of giant stars constitutes one
coherent structure with a ``core'' at $l\sim290\arcdeg$
but spreading well beyond.  This core appears to be localized
in distance.  Figure \ref{hess} shows a 2MASS Hess diagram of the Argo core,
created as the difference in Hess diagrams for fields at
$[l,b]=[285,-5]\arcdeg$ and $[l,b]=[285,+5]\arcdeg$.  From this
difference a well-defined red giant branch (RGB) associated with Argo
emerges from the ``RGB-smear'' created by closer, foreground MW disk
stars.  The RGB residue has a highest Hess diagram density that is fit
by an [Fe/H]$=-0.7$ population \citep{ivanov} at $r \sim 13.8$ kpc, but isochrones
with a spread of metallicity not unlike that previously suggested for stars in the ring
(see \S1) and at this same distance seem to reasonably account for the triangular
spread of residual red stars.  That a red clump predominantly at fainter magnitudes
remains after Hess diagram differencing is not inconsistent with the presence of
an Argo red clump at the same general distance \citep{SG02}.

Another aspect of interest is that the isodensity contours tend to have
ellipsoidal -- {\it not} warp-like --- character, with the major axes
of the isopleths ``jutting'' out obliquely from the MW disk.  In this
respect, Argo resembles a large, distorted dSph galaxy, similar to the
Sgr system.  If, like Sgr, Argo is
a coherent, symmetrical structure, we can ``fit´´ ellipses to the visible isopleths to
hypothesize its overall appearance (Fig.\ 3).  The ``eyeball´´ fit
ellipses are centered at one point in the $E_{B-V}> 0.55$ region, but have variable
axis ratio and inclination angle.  The suggestion of a ``twisting isophote''
character resembles the M giant distribution of the
 Sgr core (see, e.g., M03's Figs.\ 7e and 7f),
which, with {\it its own} ``ring'' of tidal debris wrapping around the MW, may
serve as a useful paradigm for Argo.  The
apparent inclination of Argo to the Monoceros ring (which seems to have
a closer alignment to the MW disk) recalls the fact that
the major axis of
the Sgr core is canted (by $6\arcdeg$) with respect to its tidal stream (M03). 
These structural properties of the Argo system are certainly tantalizing
support for the notion that it may be a tidally disrupting dSph-system ---
perhaps the progenitor of the Monoceros ring.


\begin{figure}
\includegraphics[angle=-90,scale=0.33]{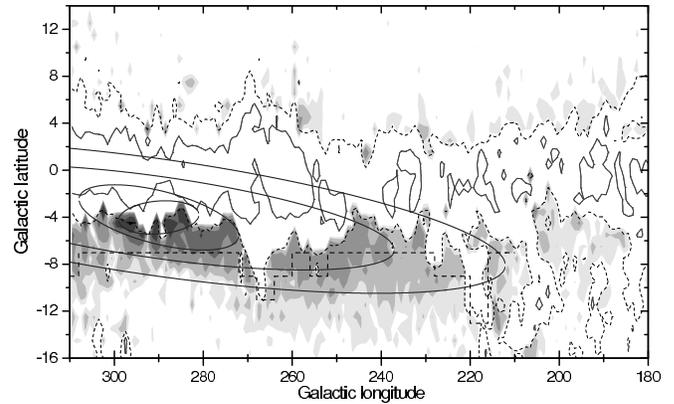}
\caption{Same as middle panel of Fig.\ 1, but with greyscales
  fit by the ellipses shown.  The dashed line indicates the $E_{B-V}=0.55$
  reddening boundary adopted in the \citeauthor*{martin04a} study.  The dotted and solid lines
  show the $E_{B-V}=0.55$ and 1.20 boundaries, respectively, from a finer
  extinction map resolution than used by \citeauthor*{martin04a}.\label{ellipses}}
\end{figure}

In their 2MASS M giant study, \citeauthor*{martin04a} also identified an apparently
ellipsoidal overdensity, but with a geometrical center at
$[l,b]\sim [240,-8]\arcdeg$.  From this core-like feature, at an
apparent distance of $10.0 \la R_{\rm GC} \la 15.8$, \citeauthor*{martin04a}
concluded that the center of the Monoceros system is in CMa.  In
contrast, our map shows nothing particularly special about this region,
except that it lies within the overall density rise towards the Argo
core, which in our map is $2-4\times$ more dense than the excess at the position of
the CMa ``core.'' Since the \citeauthor*{martin04a} analysis
is using ostensibly the same database as ours, it is critical to
understand the source of the discrepancy in results.  Clues may lie in
differences in how reddening affects each analysis:
\citeauthor*{martin04a} used both a coarser $E_{B-V}$ map resolution
{\it and} were forced to discard regions with
highly extinguished Galactic plane-mirrored counterparts that we preserve
through use of our model.  The decreased area available to the
\citeauthor*{martin04a} analysis is demonstrated by the difference
between their and our $E_{B-V}=0.55$ boundaries (Fig.\ 3).
Much of Argo, including the highest density patch, has been removed from
their analysis.  In addition, with the \citeauthor*{martin04a}
boundary it becomes less obvious that the overdensity at $l\sim240\arcdeg$
is related to the one at $l\sim290\arcdeg$ because of the intrusion of the
elongated dust feature at $l \simeq 265\arcdeg$ extending south
to $b\simeq -12\arcdeg$.  The `aproximately
ellipsoidal' shape of CMa observed by these authors may be an
artifact caused by the neighboring very reddened map cells --- the
$265\arcdeg$ reddening ``finger´´ and another at $l \simeq 220\arcdeg$; the
likely coordinates for the Monoceros structure core as given by
M04a look to be the geometrical center of the outer Argo overdensity
seen through this reddening window in their map. 
Since
the same arguments might be made about the nature and likely center of the
Argo overdensity derived from {\it our} maps, it is important to stress
both the provisional nature of our assessment of Argo and that most
important clues about this star system still remain hidden.

\begin{figure}
\includegraphics[angle=0,scale=0.5]{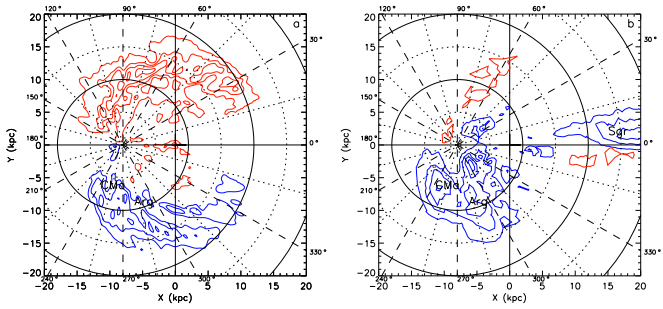}
\caption{Comparison between the gas warp and the stellar overdensities found in this paper.
The dark and light contour lines (blue and red lines, in the electronic version) in 
this plot refers to southern and northern hemisphere asymmetries, respectively,
projected onto the Galactic plane. Contour levels indicate 0.2, 0.4, 0.6 and 0.85 of the maximum gas
($5.36 M_\odot\ {\rm pc}^{-2}$) or stellar (4554 M giants ${\rm kpc}^{-1}\,{\rm rad}^{-1}$) cell density
in this plot. The position of the Sgr dSph, as well as CMa and Argo, is shown.
\label{warp}}
\end{figure}

\section{Concluding remarks}

Our analysis of the 2MASS giant star distribution in the outer Galaxy,
gives strong evidence for the presence of an
excess of stars, in the distance range of the
Monoceros ring, stretching across at least $210\arcdeg \le l \le
310\arcdeg$ with a concentration in the Argo region
($l\sim290\arcdeg$), and suggests the CMa ``core'' is a reddening 
artifact sitting on the stern of Argo. 
At minimum, our results from an improved analysis of 2MASS M giants in high $E_{B-V}$ regions 
show a different structure to the stellar density enhancements in the MW third and fourth quadrants 
than previously reported. 
Figure 3 further explores our discovered peak overdensity in Argo as if {\it it} (not CMa) 
were the core of an elongated, possibly disrupting dwarf satellite 
galaxy, perhaps the source of the Monoceros ring. But further work is necessary 
to verify this and alternative explanations.  For example, based on the distribution of 
2MASS K giant candidates, \citet{lopezcorredoira} report the discovery of a stellar warp 
in the outer disk. They base this conclusion on an apparently sinusoidal excess of stars 
with respect to $l$, under equatorial symmetry.  The south-north 
asymmetry in their data is strongest at $l\sim 245\arcdeg$, coinciding 
with CMa, but \emph{the model} proposed by the authors to reproduce these data peak 
around $l\sim 290\arcdeg$, where Argo is found.  While this may give the impression that Argo is 
the stellar component of the MW warp, their analysis  
relies on the same symmetry comparison methodology that affected other previous  
(e.g., Paper I, M04a) low latitude 2MASS studies and, as such, cannot properly 
account for the presence of any satellite near and/or spanning across the MW 
mid-plane.   

It is also useful to compare the HI Galactic warp with the stellar asymmetries 
to see whether these coincide. 
Investigations of the gas warp
show strong deviations from the Galactic plane both near $l\sim 245\arcdeg$ and
$l\sim 290\arcdeg$ (see Fig. 5 by \citealt{sodroski}). Figure \ref{warp} compares
the stellar asymmetries found by this work with the asymmetries in the gas warp,
according to the 3D data by \citet{nakanishi}. Although the southern gas warp asymmetry is similar in shape to
that coming from the stars in the third and fourth Galactic quadrants, the gas warp is 2 and
5 kpc farther from the Sun than the stellar overdensity at the longitudes of CMa and Argo, respectively. 
Moreover, quite contrary to what is seen in the stellar overdensity,
the HI warp shows a symmetrical deviation to the Northern hemisphere across the 
$l=0-180^{\circ}$ Galactic meridian and, moreover, 
it is densest in this northern hemisphere.
Since there is no counterpart to the northern hemisphere gas warp in the stellar data,
we conclude that Argo is not likely to be simply a stellar counterpart to the gas warp. 
On the other hand, it is possible that the presence of
this putative satellite galaxy could be the {\it cause} of the HI warp, 
given that
interactions with satellite dwarf galaxies is the leading explanation for disk warps
\citep{shang,schwarzkopf, reshetnikov}.

\acknowledgements

The results presented here make use of data from
the Two Micron All Sky Survey, which is a joint project of the
University of Massachusetts and the Infrared Processing and Analysis
Center, funded by the NASA and the NSF.
We acknowledge generous support in the form of a postdoctoral fellowship
for HJR-P from The F. H. Levinson Fund of the Celerity Foundation
and from NSF grant AST-0307851, NASA grant JPL 1228235 and a CNPq grant to HJR-P.

\end{document}